\documentclass[showpacs,showkeys,amsmath,amssymb,prl,tightenlines
]{revtex4-1}

\usepackage{geometry}
\usepackage{setspace}
\usepackage{dcolumn}
\usepackage{color}
\usepackage{bm}
\usepackage{graphicx}
\usepackage{epsfig}
\usepackage{amsmath}
\usepackage{amsfonts}
\usepackage{amssymb}
\newcommand{\beq}{\begin{equation}}
\newcommand{\eeq}{\end{equation}}
\newcommand{\eqna}{\begin{eqnarray}}
\newcommand{\eqne}{\end{eqnarray}}
\global \def \eqa#1 {\begin{align} #1 \end{align}}
\global \def \eqan#1 {\begin{align*} #1 \end{align*}}
\newcommand{\beqa}{\begin{align*}}
\newcommand{\eeqa}{\end{align*}}

\newcommand{\vecr}{\mathbf{r}}

\newcommand{\ha}{\hat{a}}

\newcommand{\la}{\langle}
\newcommand{\ra}{\rangle}

\begin{document}
\title{
\bf {Variational Approach to Tunneling Dynamics. \\ Application to  Hot Superfluid Fermi Systems. \\ Spontaneous and Induced Fission.} }
\date{\today}
\author{  S. Levit}
\affiliation{ Department of Condensed Matter Physics \\  Weizmann Institute of Science \\ Rehovot 761001, Israel}
\email{shimon.levit@weizmann.ac.il}
\begin{abstract}

We introduce a general variational framework to address the tunneling of hot Fermi systems. We use the representation of the trace of the imaginary time $\tau=it$ propagator as a functional integral type of a sum over complete sets of states at intermediate propagation slices. We assume that these states are $\tau$-dependent and generated by an arbitrary  trial Hamiltonian $H_0(\tau)$.  
We then use the convexity inequality to derive $H_0(\tau)$ controlled variational bound  for a trial action functional. This functional has a general structure consisting of two parts -  statistically weighted quantum penetrability  and dynamical tunneling entropy. We examine how this structure incorporates the basic physics of tunneling  of hot Fermi systems. Using the  variational inequality one can optimise the dynamical parameters controlling the action functional for any choice of the trial problem. 

As an application we take $H_0(\tau)$ to describe imaginary time dynamics of non interacting  Bogoliubov-de Gennes (BdG) quasiparticles. Optimising its dynamical parameters we extend the tunneling  theory of hot Fermi systems to the Hartree-Fock-Bogoliubov(HFB) frame and derive the corresponding generalisation of imaginary time temperature dependent  BdG mean field equations.   

As in the trial action the  prominent feature of these equations is an inseparable interplay between quantum dynamical and entropic statistical effects.  In the zero temperature limit these  equations describe the "false ground state" tunneling decay of superfluid Fermi systems (spontaneous fission in nuclear physics).  With increasing excitation energy (effective temperature) the decay process is gradually evolving from pure quantum tunneling to statistical "bottle neck" escape mechanism.  Correspondingly the statistically weighted dynamical penetrability part in the  action gradually decreases while the tunneling entropy increases with  increasing effective temperature.

 \end{abstract}
\maketitle
\section{Motivation. Statistically averaged penetrability and tunneling entropy.}

Continuation of classical equations to imaginary time domain have long been understood as a way to describe quantum mechanical tunneling phenomena, cf., Ref.\cite{Col}.  By extending the field theoretical description of "false vacuum decay", Ref.\cite{ColCal},  the fermion mean field tunneling description of spontaneous fission, i.e. decay of a "false ground state" was proposed in Refs.\cite{LevNeg,Rein1}.  This formalism essentially amounts to imaginary time  $\tau = it$ dependent Hartree-Fock (ITDHF) equations \cite{Neg}. The tunneling bounce solution appears as $\tau$ dependent Slater determinant eigenfunction of a boundary problem with periodic boundary conditions on the imaginary time interval of asymptotically infinite extent  and single particle quasi-energies as eigenvalues. Results of application of this theory to model problems were reported in Refs. \cite{LevNeg, Rein2, Pud, Arve, Neg}. Algorithmic challenges to find periodic mean field solutions were discussed in Ref.\cite{Bar}.
 
Extension of the  mean field equations of "false ground state decay" was proposed in Ref.\cite{KerLev} to describe tunneling decay of a micro-canonical ensemble of excited states of an interacting self-bound Fermi system with fixed excitation energy $E^*$ and particle number $N$.  Such decays provide a common description of the so called induced  fission, cf., Ref.\cite{SchRob}. The resulting dynamical equations describe statistically  averaged tunneling bounce of Slater determinants.  This theory was based on auxiliary field functional integral formalism with the resulting equations of the Hartree type. The ways to extend this to the Hartree-Fock frame were outlined in Refs.\cite{JPBOrl} and \cite{KerLevTro}.

Here we introduce a more general variational approach to imaginary time dynamics of equilibrated Fermi systems. We use the representation of the trace of the imaginary time propagator 
$Tr e^{-\beta(\hat{H} - \mu \hat{N})}$  as a functional integral type of a sum over complete sets of states at the intermediate propagation slices. We assume that these states are $\tau$-dependent and generated by a trial Hamiltonian $H_0(\tau)$. Apart of the symmetry condition $H_0(\tau) = H_0(-\tau)$ for $-\beta/2 \le \tau \le \beta/2$  this Hamiltonian can be  arbitrary.  We then use the convexity inequality to derive a variational bound   for the trial action functional which has the following form
\beq \label{action_1}
-\ln Tr e^{-\beta(\hat{H} - \mu \hat{N})} \le \sum_K W_K^{(0)} \int_{-\beta/2}^{\beta/2} d\tau \langle \bar \Psi^{(0)} _K(\tau)|\frac{\partial}{\partial \tau}  + \hat{H} -\mu \hat{N}| \Psi_K^{(0)} (\tau)\rangle  + \sum_K W_K^{(0)}\ln W_K^{(0)} 
\eeq
where $\hat{H}$ is the exact Hamiltonian, $|\Psi_K^{(0)} (\tau)\rangle$'s are solutions of the quasienergy problem defined by the trial $\hat{H}_0(\tau)$  on the interval $-\beta/2 \ge \tau \ge \beta/2$, cf., Eq.(\ref{bc_problem_trial_in_tau_0})  below and $W_K^{(0)}$ are statistical weights defined in terms of the quasienergies $\Lambda_K$, cf., Eqs.(\ref{W_Ks} , \ref{W_Ks_first}).  

The first part of the functional on the  right hand side of  (\ref{action_1}) has the form of statistically averaged action while the second part can be interpreted as dynamical entropy. Using the  above inequality one can optimise the parameters of the trial  $\hat{H}_0(\tau)$ controlling this  functional for a given $\beta$ and $\mu$ and any choice of $\hat{H}_0(\tau)$ symmetric in $\tau$.  In the application below we will illustrate this procedure and the use of its results to estimate the tunneling probability, cf., Eq. (\ref{tun_prob}).

In order to understand the physics of the trial action in  (\ref{action_1}) let us consider tunneling of a hot Fermi system like e.g. induced fission of an equilibrated  (compound) nucleus having fixed  excitation energy $E^*$ and a given particle number $N$.  
Such a system must be described by a micro-canonical ensemble which means that the values of $\beta$ and $\mu$ will be fixed by $E^*$ and $N$. After the optimization the terms with $\hat{H}$ and $\hat{N}$ in the trial action will be removed by the appropriate Legendre transformations or alternatively by the saddle point approximation to the integral relation between the grand canonical and the micro-canonical formulations.  
As a result the first part of the trial action in  Eq. (\ref{action_1}) would consist of only the $\partial/\partial \tau$ term which should clearly be viewed as statistically weighted quantum penetrability.  The second part has the form of the statistical entropy which however depends on the tunneling dynamics via quasi-energies. We shall call it tunneling entropy.  The first term vanishes for a static trial problem while the last tends to zero at zero temperature, $\beta \to \infty$   with the statistical weights becoming stepwise $1$ to $0$ function.   

To understand the role and the interplay of the averaged penetrability and tunneling entropy let us consider the following schematic semiclassical picture of tunneling of such systems. Imagine a fly caught in a bottle with a narrow opening on the side. The bottle is the phase space of an (equilibrated) system (all its 6N degrees of freedom). The fly is the point representing the state of the system in the ensemble.   At $E^*=0$ the phase space is just one point and the only way for the fly to escape is via tunneling through the bottle's ''thick walls''.  This is the spontaneous fission, $\beta \to \infty$ limit.

 As $E^*$ goes up the  phase space for the fly grows. It can still tunnel from the points where the bottle wall is the thinnest (at this $E^*$) i.e. have the largest quantum tunneling probability or from other places (phase space volume). The wall is thicker there (smaller tunneling probability)  but there are many more such points (larger entropy). So the tunneling becomes a mixture of QM tunneling and entropic effects. 

The important point is that even as $E^*$ is above the bottle opening (i.e. above the $T=0$ barrier energy) the same competition is going on - the fly can either go through the opening (where no quantum tunneling is needed) from a small number (volume) of points  ''near the opening'' or tunnel from other, much larger in number (high entropy) points but pay the price of quantum tunneling.  

This is consistent with the recent study \cite{KerLate} of the excitation energy $E^*$  dependence of isentropic fission barriers. The barrier height $E_B(E^*)$ is not a fixed (zero temperature) parameter. As  $E^*$  grows $E_ B(E^*)$ grows with it and although the difference $E_ B(E^*) - E^*$ decreases it vanishes only asymptotically at large values of $E^*$. 

As a first application of our formalism we derive the generalisation of the Bogoliubov-de Gennes (BdG) mean field equations to describe imaginary time tunneling bounce at zero and finite effective temperatures/excitation energies. Such a generalisation is needed since significant changes of the mean field shape during tunneling lead to multiple crossings of single particle levels with different symmetry, Ref. \cite{Pud,Neg}. Effective switching between such levels requires presence of terms in the mean field Hamiltonian which are absent in the Hartree-Fock approximation.  Theoretical considerations  supported by growing body of numerical simulations clearly indicate that  the pairing interaction is often the missing component in the HF theories of collective tunneling dynamics \cite{BriBro} - \cite{SudDob}. The Hartree-Fock-Bogoliubov (HFB) extension appears to be the most suitable framework. 

Constraining $H_0(\tau)$ to have the HFB form describing non interacting Bogoliubov-de Gennes (BdG) quasiparticles we optimise
 the dynamical parameters of $H_0(\tau)$.  As a result  we derive an extension of the BdG equations  with both  imaginary time and temperature dependent mean field.  The quasiparticle eigenfunctions are bouncing in the imaginary time inverse temperature interval while the self-consistent density and pairing matrices depend on the thermal Fermi occupations which in turn depend on the quasiparticle quasienergies, cf., Eq.(\ref{HFB_sp_modes_1}).  Thus as in the trial action an inseparable interplay exists between quantum dynamical and entropic statistical effects.  With increasing excitation energy (effective temperature) the decay process is gradually evolving from pure quantum tunneling to statistical "bottle neck" escape mechanism.  Correspondingly the statistically weighted dynamical penetrability part in the  action gradually decreases while the tunneling entropy increases with  increasing effective temperature. 
 
 Theory of nuclear fission is a self evident field of application of our development. Impressive recent improvements of numerical techniques and computer resources should help to apply our results for improving the microscopic description of spontaneous and low energy induced fission. It should also be useful in providing  novel microscopic insights in the existing phenomenological theories of the fission phenomena.

 \section{Imaginary time propagation in an arbitrary many body basis}
Assume many  fermion Hamiltonian  \beq \label{many_body_H}
\hat{H} =   \hat{T} + \hat{V} =
 \sum_{ij} t_{ij} \ha^+_i \ha_j  \; + \;  \frac{1}{2} \sum_{ijkl} V_{ijkl} \ha^+_i\ha^+_j \ha_l \ha_k   
\eeq
in terms of standard  Fermi operators $\ha_i$ and $\ha_j^+$ in an arbitrary single particle basis $\{\phi_k(\vecr\sigma)\equiv \phi_k(x)\}$ with one body part  $t_{ij}$ representing kinetic energy and possibly external potential and two body interaction $V_{ijkl}$. 

The most common Fermi systems which exhibit tunneling are heavy nuclei undergoing fission. The most common initial state is the so called compound nucleus which can statistically be described as an approximately equilibrated system with a given excitation energy and a particle number. Formally this is a microcanonical ensemble with the partition function $Tr\delta(E- \hat{H} ) \delta (N - \hat{N})$. In a standard way it can be related to the grand canonical partition function 
\beq \label{exact_Z_0}
Z(T,\mu) =Tr e^{-\beta(\hat{H} - \mu \hat{N})} = \sum_{K} \langle \Psi_K | e^{-\beta(\hat{H} - \mu \hat{N}) }| \Psi_K \rangle \;\;\;, \;\;\; \beta = 1/T
\eeq
where $\{\Psi_K\}$ denotes a complete set of many fermion states in the Fock space of the system and $\beta$ and $\mu$ are determined in a standard way by $E$ and $N$, cf., Ref.\cite{KerLev}.

Let us slice the exponential under the trace in (\ref{exact_Z_0})
\beq \label{imag_time_slices}
Tr e^{-\beta(\hat{H} - \mu \hat{N})} = \lim_{\epsilon \to 0} \;  \sum_{K} \langle \Psi_K | \prod_{m=1}^M [1-\beta\epsilon (\hat{H} - \mu \hat{N}) ] | \Psi_K \rangle  \;\;\; , \;\;\; \epsilon=\frac{1}{M}
\eeq
and insert complete sets of (many fermion) states  at each slice in the expression (\ref{imag_time_slices}). We label these sets by imaginary time $\{\Psi_K (\tau_m) \}$ with $-\beta/2\le \tau \le \beta/2 $ and moreover choose them not identical but changing from slice to slice and take them obeying
\beq \label{def_of_Psi_bar}
\{\Psi_K (\tau_m) \}= \{\Psi_K (- \tau_m)\} \;\; ; \;
 \{\Psi^*_K (\tau_m) \} = \{\Psi^*_K (-\tau_m) \} \equiv \{\bar\Psi_K (\tau_m) \}
\eeq
In a standard way we obtain in the limit $M\to\infty$ a functional integral like sum 
\eqa{ \label{exact_Z_3}
Z(T,\mu)= \sum_{[\Psi^{(0)}_K(\tau)]} \exp\left\{-\int_{-\beta/2}^{\beta/2} d\tau \left[\langle \bar{\Psi}^{(0)}_K (\tau) |\frac{\partial}{\partial \tau}  + \hat{H}  - \mu \hat{N} | \Psi^{(0)}_K(\tau) \rangle \right]  
\right\} 
}
where the condition (\ref{def_of_Psi_bar}) assures that each term in this sum is real and positive.

There are several ways to proceed from this exact expression to devise a suitable mean field theory.  These include the superfluid density functional approach or the use of functional integral techniques. We will describe them elsewhere, Ref.\cite{LevPrep}. Here we will follow a variational approach.

\section{Variational framework - using the convexity inequality}

Let us assume some trial Hamiltonian $\hat{H}_0(\tau)$ which is in general $\tau$ dependent. At the moment there is no need to specify the exact form of $\hat{H}_0(\tau)$ apart of it being symmetric $\hat{H}_0(\tau) = \hat{H}_0(-\tau)$.  In the following we will consider a specific example of $\hat{H}_0(\tau)$ leading to finite temperature imaginary time dependent Hartree-Fock-Bogoliubov(HFB) mean field formalism. 

 Let us add and subtract $\hat{H}_0(\tau)$ in the exponential of (\ref{exact_Z_3}) and write the result as
 \eqa{ \label{exact_part}
Z(T,\mu) \equiv Tr e^{-\beta(\hat{H} - \mu \hat{N})} & =   Z_0(T,\mu)\sum_{[\Psi_K(\tau)]}  W^{(0)}_K \exp \left(-\int_{-\beta/2}^{\beta/2} d\tau \left[ \langle \bar{\Psi}_K(\tau) |  \hat{H}  - \hat{H}_0(\tau) | \Psi_K(\tau) \rangle \right] \right)    
}
where we defined the corresponding probabilities 
 \beq \label{W_Ks_first}
W^{(0)}_K = \frac{1}{Z_0(T,\mu)} \exp \left\{-\int_{-\beta/2}^{\beta/2} d\tau \langle \bar{\Psi}_K (\tau) | \frac{\partial}{\partial \tau}  + \hat{H}_0(\tau)  - \mu \hat{ N} | \Psi_K(\tau) \rangle \right\}
\eeq
The symmetry $\hat{H}_0(\tau) = \hat{H}_0(-\tau)$ and Eq.(\ref{def_of_Psi_bar}) assure $W^{(0)}_K\ge 0$ and $Z_0(T,\mu)$ normalises $\sum_K W^{(0)}_K =1$.

Using $\langle e^{-X}\rangle \ge e^{-\langle X\rangle}$, Ref.\cite{ineq}, in Eq.(\ref{exact_part})  we arrive at a variational inequality 
\beq \label{Omtrial0}
\Omega(T,\mu) =-T\ln Z(T,\mu) \le \Omega_{trial}(T,\mu) = \Omega_0(T,\mu) +\frac{1}{\beta} \int_{-\beta/2}^{\beta/2} d\tau \langle \hat{H} -
 \hat{H}_0(\tau) \rangle 
\eeq
with the notation $\Omega_0(T,\mu)   = -T \ln     Z_0(T,\mu)$  and 
$\la \hat O(\tau)\ra =\sum_K W_K^{(0)} 
 \langle \bar \Psi^{(0)}_K(\tau) |  \hat O(\tau)  | \Psi^{(0)}_K(\tau) \rangle $.  
 
To evaluate $\Omega_{trial}(T,\mu)$ it is useful to introduce the imaginary time evolution operator for $\hat{H}_0(\tau) - \mu\hat{N}$ 
\beq \label{evol_op}
\hat{U}_0(\tau,-\beta/2) = T_\tau \exp\{-\int_{-\beta/2}^\tau d\tau' [ \hat{H}_0(\tau') - \mu\hat{N}] \}
\eeq
It is straightforward  to show via time slicing  that  
\beq \label{Omega0_via_U}
 Z_0(T,\mu) =  Tr \, \hat{U}_0(\beta/2 , -\beta/2)\;\;\; \to \;\;\; \Omega_0(T,\mu) = -T \ln Tr  \, \hat{U}_0(\beta/2 , -\beta/2) 
\eeq
 Let us  consider the eigenvalue equation
\beq \label{eigen_func_of_U}
\hat{U}_0(\beta/2,-\beta/2)|\Phi_K\rangle = e^{-\beta \Lambda_K} |\Phi_K \rangle
\eeq
The operator $\hat{U}_0(\beta/2,-\beta/2)$ is Hermitian and moreover positive for the symmetric 
$\hat{H}_0(\tau) =  \hat{H}_0(-\tau)$ so one can  choose $\Lambda_K$ as real.  We shall refer to them as quasienergies.  We can write
\beq \label{W_Ks}
Z_0 = \sum_K e^{-\beta \Lambda_K}  \;\;\;\; \to  \;\;\;\;  W_K^{(0)} = \frac{1}{Z_0} e^{-\beta\Lambda_K}
\eeq
A practical way to work with $\hat{U}_0(\tau,-\beta/2)$ is to introduce the $\tau$ dependent
\beq
| \Phi_K(\tau)\rangle \equiv \hat{U}_0(\tau,-\beta/2) | \Phi_K \rangle
\eeq
in terms of which the eigenvalue equation (\ref{eigen_func_of_U}) is equivalent to solving the boundary value problem 
\beq \label{bc_problem_for_PSI}
[ \frac{\partial}{\partial \tau} +  \hat{H}_0(\tau)-\mu\hat{N}] |\Phi_K(\tau)\rangle =0  \;\; , \;\;   |\Phi_K(\beta/2)\rangle = e^{-\beta\Lambda_K} 
|\Phi_K(-\beta/2)\rangle
\eeq
 It is furthermore convenient to introduce $|\Psi_K(\tau)\rangle = e^{\Lambda_K (\tau +\beta/2)} |\Phi_K(\tau)\rangle$  which satisfies 
\beq \label{bc_problem_trial_in_tau_0}
    [ \frac{\partial}{\partial \tau} +  \hat{H}_0(\tau) - \mu \hat{N}] |\Psi_K^{(0)}(\tau)\rangle =     \Lambda_K   |\Psi^{(0)}(\tau)\rangle\;\;\; {\textrm with}  \;\;\;   |\Psi_K^{(0)}(\beta/2)\rangle =    |\Psi_K^{(0)}(-\beta/2)\rangle
 \eeq 
 These equations together with the identity $-\ln  Z_0 = \beta \sum_K \Lambda_K W_K^{(0)} + \sum_K W_K^{(0)} \ln W_K^{(0)}$ allow to write 
 \beq
\beta \Omega_0 = \sum_K W_K^{(0)} \int_{-\beta/2}^{\beta/2} d\tau \langle \bar \Psi^{(0)} _K(\tau)|\frac{\partial}{\partial \tau}  + \hat{H}_0 -\mu \hat{N}| \Psi_K^{(0)} (\tau)\rangle  + \sum_K W_K^{(0)}\ln W_K^{(0)} 
 \eeq
 Inserting this into the inequality (\ref{Omtrial0}) allows to write it in a form which was quoted in the Introduction, Eq. (\ref{action_1}). 
 
 As a last remark let us indicate that the transformation from the equations for $ |\Phi_K(\tau)\rangle$ to the boundary problem (\ref{bc_problem_trial_in_tau_0}) introduces (gauge like) ambiguities in the definitions of $ \Lambda_K$ and   $ |\Psi^{(0)}(\tau)\rangle$. We refer the reader to Ref.\cite{Pud} for the discussion of how to deal with this issue.
 
\section{Thermal tunneling bounce -  imaginary time dependent Bogoliubov-de Gennes equations}
We take the trial Hamiltonian  $\hat{H}_0$ to have the form 
\beq \label{trial_HFB_H0}
\hat{H}_0(\tau)=\sum_{ij}\left [ (t_{ij} +\gamma_{ij} (\tau))  \ha^+_i\ha_j + \frac{1}{2} \Delta_{ij}(\tau) \ha^+_i\ha^+_j + \frac{1}{2} \Delta^*_{ij}(\tau) \ha_i\ha_j \right]
\eeq
with   
\beq \label{gamma_Delta_in_sigma_eta}
\gamma_{ij}(\tau)=  \sum_{kl}V^{A}_{ikjl}\, \sigma_{lk}(\tau) \; \; \; , \; \; \;\Delta_{ij} (\tau) =\sum_{kl} V_{ijkl}\, \eta_{lk}(\tau)
\eeq
and  $\sigma(\tau)$ and $\eta(\tau)$ matrices obeying $\sigma_{ij}(\tau) =  \sigma^*_{ji}(\tau)  ,  \eta_{ij}(\tau) = -\eta_{ji}(\tau)$ considered as variational parameters. 
  
   One needs to solve Eq.(\ref{bc_problem_trial_in_tau_0}) with $H_0(\tau)$ given by Eq.(\ref{trial_HFB_H0}), use the solutions  to express 
$\Omega_{trial}(T,\mu)$ of Eq. (\ref{Omtrial0}) as a functional of $\sigma_{ij}(\tau)$'s  and $\eta_{ij}(\tau)$'s and then minimise it.  One can do this generalising the derivation of the static finite temperature Hartree-Fock-Bogoliubov formalism. The details will be published elsewhere \cite{LevPrep}. Here is the summary.

Let us define ($\tau$ - dependent) quasiparticle operators 
\beq \label{def_of_qp_op}
 \hat{d}^{+}_\nu (\tau)= \sum_i \left[u_{\nu} (i,\tau)\hat{a}^{+}_i + v_{\nu}(i,\tau)  \hat{a}_i\right]
 \;\;\; , \;\;\;
\hat{\bar d}_\nu(\tau) = \sum_i \left[\bar u_{\nu} (i,\tau)\hat{a}_i + \bar v_{\nu} (i,\tau) \hat{a}^{+}_i\right] 
\eeq
where $\bar u_\nu(\tau) \equiv u^*_{\nu} (-\tau)\; , \; \bar v_\nu(\tau) \equiv v^*_{\nu} (-\tau)$. The functions
$u_{\nu} (\tau) , v_{\nu} (\tau)$ satisfy the eigenvalue equations 
\eqa{ \label{HFB_sp_modes}
\left(  \frac{\partial}{\partial \tau} +  \mathcal{H} (\tau) \right)  w_\nu(\tau)  =  \lambda_\nu w_\nu(\tau)   \;\;\;\;\;\;   ,   \;\;\;\;\; w_\nu(\beta/2) = w_\nu(-\beta/2) 
}
 where denoted
 \eqan{
 \mathcal{H} (\tau) = \left(
\begin{array}{cc}  h(\tau) -\mu & \Delta(\tau)  \\   - \Delta^*(\tau)   &  -h^*(\tau)  +\mu  \end{array} \right)  \;\;\;\;\; , \;\;\;\;\;\;
w_\nu(\tau)  =  \left( \begin{array}{c} u_\nu(\tau) \\ v_\nu (\tau)   \end{array} \right) 
}
With such $u_{\nu} (\tau) , v_{\nu} (\tau)$ the operators $\hat{\bar d}_\nu(\tau) ,  \hat{d}^{+}_\nu (\tau)$ satisfy the Fermi commutation relations and one can write the solutions of  Eq.(\ref{bc_problem_trial_in_tau_0}) as 
\beq\label{HFB_wavefunc}
|\Psi_K(\tau)\rangle \equiv   |\Psi_{n_1,...,n_\nu, ....} (\tau)\rangle =   \prod_{\nu > 0}  [\hat{d}^+_{\nu} (\tau)]^{n_\nu}  |\Psi_0(\tau) \rangle \;\;\;  {\rm with} \;\;   n_\nu = 0 \; {\rm or} \; 1
\eeq
 This provided that the quasiparticle vacuum $|\Psi_0(\tau) \rangle$ is annihilated by all $\hat{\bar d}_\nu(\tau)$'s. The corresponding  $\Lambda_K=\Lambda_0 + \sum_\nu\lambda_\nu$.
 
 Using such $|\Psi_K(\tau)\rangle$'s in $\Omega_{trial}(T,\mu)$, Eq.(\ref{Omtrial0}), and minimising the result with respect to $\sigma_{ij}(\tau)$'s  and $\eta_{ij}(\tau)$'s leads to the self consistency conditions
\eqa{\label{HFB_sp_modes_1}
\sigma_{ji}(\tau) & = \sum_{\nu>0} [ (1-f_\nu) \bar v_\nu (j, \tau)  v_\nu(i ,\tau)   + f_\nu   u_\nu (j,\tau)  \bar u_\nu ( i, \tau) ] 
\nonumber \\
 \eta_{ji} (\tau) &=  \sum_{\nu>0} [ (1-f_\nu)   u_\nu (i,\tau)  \bar v_\nu (j, \tau)  + 
  f_\nu \bar v_\nu (i, \tau)  u_\nu (j,\tau)   ]  \nonumber \\
  f_\nu &= \frac{1}{1+e^{\lambda_\nu}}
}
It is instructive to examine several limits of the above equations. In the static, $\tau$ independent limit one readily finds the standard, temperature dependent HFB expressions. Disregarding the pairing field $\Delta$ in the trial Hamiltonian (\ref{trial_HFB_H0}) leads to the imaginary time dependent HF form of the equations with thermal averaged mean field. These were first derived in the Hartree form in Ref.\cite{KerLev} where it was argued that they provide a mean field description of tunneling in induced fission.  At zero temperature the HF limit of the above equations reduce to the HF equations describing spontaneous  fission, cf., Refs.\cite{LevNeg, Rein1, Pud, Neg}. The zero  temperature limit of our equations provide the  HFB generalisation of that theory.  

Solution of the equations  Eqs. (\ref{HFB_sp_modes} - \ref{HFB_sp_modes_1}) generate an (approximate)  contribution to the partition function given by the exponential of $-\beta\Omega_{trial}(T,\mu)$ as given by Eq.(\ref{Omtrial0}). As was outlined in the discussion after 
Eq.(\ref{action_1}) the contribution of such solution to  microcanonical partition function is simply obtained by dropping the terms $\hat{H}(\tau) -\mu \hat{N}$ from the expression for $\Omega_{trial}$ and expressing the inverse temperature $\beta$ and chemical potential $\mu$ via the excitation energy $E^*$ and the particle number $N$. 

Taking into account the contribution from the static mean field one can follow Ref.\cite{KerLev}  and write for the tunneling decay width (inverse of the tunneling probability per unit time) 
\beq \label{tun_prob}
\Gamma(E^*,A) = \frac{D(E^*,A)}{2\pi} e^{ - \mathcal{R}(E^*,N) +\mathcal{S}(E^*,N)}
\eeq
with 
\beq \label{tun_ent_prob}
   \mathcal{R} =  \sum_K W_K^{(0)} \int_{-\beta/2}^{\beta/2} d\tau \langle \bar \Psi^{(0)} _K(\tau)|\frac{\partial}{\partial \tau} | \Psi_K^{(0)} (\tau)\rangle   \;\;\;, \;\;\; \mathcal{S} = - \sum_K W_K^{(0)}\ln W_K^{(0)} =   -\sum_\nu [ f_\nu \ln f_\nu + (1-f_\nu) \ln(1-f_\nu)]
\eeq
and $D(E^*,N) = \rho(E^*,N)^{-1}$ - the inverse level density of the decaying system (compound nucleus).   Both terms in the exponential have statistical (via $W_K^{(0)}$'s) and imaginary time dynamic (via Eqs. (\ref{HFB_sp_modes} - \ref{HFB_sp_modes_1})) contents.

{99} 
\end{document}